\documentclass{jfm}

\usepackage{graphicx}
\usepackage{natbib}
\usepackage{epstopdf}
\usepackage{amsmath}
\usepackage{esdiff} 
\usepackage{subfig}
\usepackage{float}

\ifCUPmtlplainloaded \else
 \checkfont{eurm10}
 \iffontfound
 \IfFileExists{upmath.sty}
 {\typeout{^^JFound AMS Euler Roman fonts on the system,
 using the 'upmath' package.^^J}%
 \usepackage{upmath}}
 {\typeout{^^JFound AMS Euler Roman fonts on the system, but you
 dont seem to have the}%
 \typeout{'upmath' package installed. JFM.cls can take advantage
 of these fonts,^^Jif you use 'upmath' package.^^J}%
 }
 \else
 \fi
\fi

\ifCUPmtlplainloaded \else
 \checkfont{msam10}
 \iffontfound
 \IfFileExists{amssymb.sty}
 {\typeout{^^JFound AMS Symbol fonts on the system, using the
 'amssymb' package.^^J}%
 \usepackage{amssymb}%
 \let\le=\leqslant \let\leq=\leqslant
  
 }{}
 \fi
\fi

\ifCUPmtlplainloaded \else
 \IfFileExists{amsbsy.sty}
 {\typeout{^^JFound the 'amsbsy' package on the system, using it.^^J}%
 \usepackage{amsbsy}}
 {\providecommand\boldsymbol[1]{\mbox{\boldmath $##1$}}}
\fi

\newsavebox{\astrutbox}
\sbox{\astrutbox}{\rule[-5pt]{0pt}{20pt}}

\title{Forced motion of a cylinder within a liquid-filled elastic tube}
\shorttitle{Forced motion of a cylinder within an elastic tube}
\author[Amit Vurgaft, Shai B. Elbaz and Amir D. Gat]{Amit Vurgaft, Shai B. Elbaz and Amir D. Gat}
\affiliation{Faculty of Mechanical Engineering, Technion - Israel Institute of Technology, Haifa 3200003, Israel}

\pubyear{2018} \volume{999} \pagerange{999--9999}
\date{2019}
\begin{document}
\maketitle

\abstract{This work analyzes the viscous flow and elastic deformation created by the forced axial motion of a rigid cylinder within an elastic liquid-filled tube.  The examined configuration is relevant to various minimally invasive medical procedures in which slender devices are inserted into fluid-filled biological vessels, such as percutaneous revascularization, interventional radiology, endoscopies and catheterization. By applying the lubrication approximation, thin shell elastic model, as well as scaling analysis and regular and singular asymptotic schemes, the problem is examined for small and large deformation limits (relative to the gap between the cylinder and the tube). At the limit of large deformations, forced insertion of the cylinder is shown to involve three distinct regimes and time-scales: (i) initial shear dominant regime, (ii) intermediate regime of dominant fluidic pressure and a propagating viscous-peeling front, (iii) late-time quasi-steady flow regime of the fully peeled tube. A uniform solution for all regimes is presented for a suddenly applied constant force, showing initial deceleration and then acceleration of the inserted cylinder. For the case of forced extraction of the cylinder from the tube, the negative gauge pressure reduces the gap between the cylinder and the tube, increasing viscous resistance or creating friction due to contact of the tube and cylinder. Matched asymptotic schemes are used to calculate the dynamics of the near-contact and contact limits. We find that the cylinder exits the tube in a finite time for sufficiently small or large forces. However, for an intermediate range of forces the radial contact creates a steady locking of the cylinder inside the tube.}
 
\section{Introduction} 
This work studies the dynamic response of a liquid-filled tube due to the forced axial motion of an internal rigid cylinder. This configuration is relevant to various minimally invasive medical procedures in which slender devices are inserted into fluid-filled biological vessels. For example, recent technologies for percutaneous revascularization involve insertion of cylindrical devices into blocked blood vessels \citep{rogers2007overview, davis2015no}. Similar methods are used in the field of interventional radiology, such as laser angioplasty \citep{serruys1993quantitative}, microvascular plug deployment \citep{pellerin2014microvascular} and removing blood clots by thrombolysis and thrombectomy \citep{dunn2015thrombectomy}. Additional relevant procedures are endoscopies of body organs which contain liquid, e.g. cystoscopy \citep{chew1996urethroscopy}, as well as the frequently used procedure of urinary catheterization \citep{nacey1993evolution}.

The examined configuration is actuated by an external force which induces a viscous flow-field, applying fluidic stress on the fluid-solid interface and creating deformation of the tube, thus modifying the flow-field. This interaction between viscous and elastic effects is relevant to various research fields \citep{duprat2015fluid}, including locomotion at low-Reynolds-numbers \citep{wiggins1998flexive,camalet2000generic}, flow in flexible and collapsible tubes \citep{heil1996stability,heil1998stokes, marzo2005three} and the dynamics of membrane-bound particles \citep{vlahovska2011dynamics, abreu2014fluid} among many others \citep{lister2013viscous,hewitt2015elastic,elbaz2016axial}. 

Fluid-solid interactions, in geometries similar to the one investigated here, have been extensively studied in the context of medical operations and biological flows. For example, previous works analyzing the fluid mechanics of catheterized arteries include \cite{karahalios1990some}, who investigated flow in an axisymmetric cross-section of a catheterized artery, and estimated the shear stress at the artery wall due to catheterization. \cite{sarkar2001nonlinear} modeled the pulsating blood flow in the annular cross-section between a catheter and an elastic tube and calculated the induced pressure gradient along the elastic tube. \cite{vajravelu2011mathematical}  modelled a non-Newtonian Herschel-Bulkley fluid flow in an elastic tube, representing a catheterized artery. \cite{kumar} showed that the effective viscosity, flow rate and arterial wall shear stress are significantly altered in the catheterized site.

Other relevant works studied the motion of closely-fitting solids in elastic tubes filled with viscous fluid, as a model of blood cells in narrow capillaries. Such problems involve the effects of the hydrodynamic stress generated in the lubricating film between the two bodies and the elastic stress which develops as a result of the contact of the particle and the tube. This type of analysis was first done by \cite{lighthill1968pressure} who provided analytical solutions for the pressure-field in such configurations, as well as predicting a necking phenomena next to the contact point. Later, \cite{tozeren1982flow} found the force required to maintain the motion of the particle, in addition to calculation of the fluid pressure-field and the elastic deformation. \cite{tani2017motion} examined the friction force between the inner solid and the elastic tube in the case of a dry sphere-tube contact and in the case of lubrication by a thin fluid layer. Another recent relevant work is \cite{park2018viscous}, who presented both analysis and experimental data of viscous flow in a bio-inspired soft valve configuration. In this context, the authors studied a cylinder and a concentric sphere with a narrow gap between them, and obtained the effect of the sphere on the nonlinear relation between the externally  applied pressure difference and the flow rate. 

The aim of this work is to analyze the motion of a solid cylinder within a liquid-filled tube, due to a prescribed external force, in relevance to various medical procedures. In \S2 we begin by defining the geometrical and physical properties of the examined configuration, as well as stating the governing lubrication equation and integral constraints of the system.  Analysis of the simpler case of small linearized elastic deformations is presented in \S3.  We then present the large deformation nonlinear dynamics for insertion (\S4)  and extraction (\S5) of the cylinder from the elastic tube. In \S6 we provide concluding remarks. 

\section{Problem formulation \& governing equations}
\begin{figure}
\centering
\includegraphics[width=1\textwidth]{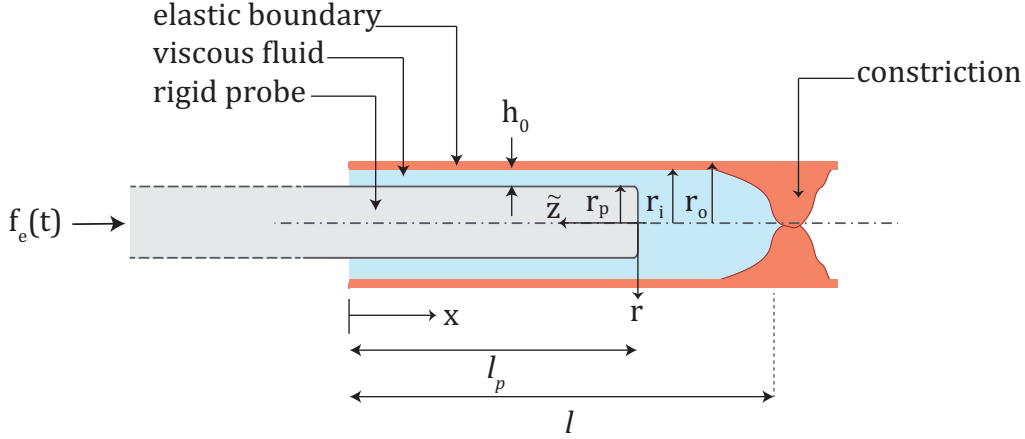}
\captionsetup{justification=justified}
\caption{\textbf{Illustration of the examined configuration (at rest) and definition of the coordinate system.}  A rigid tube is located within a semi-closed liquid-filled elastic tube. The cylinder can be extracted from, or inserted into, the tube by an external axial force. $h_0$ is the initial gap between the cylinder and the tube, $l_p$ is the length of the inserted tube, $r_p$ is the radius of the tube, $l$ is the length from the inlet to the constriction closing the tube, $r_i$ and $r_o$  are the inner and outer radii of the tube, respectively.}
\label{figure_extension}
\end{figure}

We study a Newtonian, incompressible, creeping flow due to the motion of a rigid cylinder within a liquid-filled elastic tube, as shown in figure \ref{figure_extension}. The inner cylinder represents a simplified model of a minimally invasive medical device and the linearly elastic tube is a simplified model of a bounding biological vessel (artery, urethra, etc.). An external force is applied on the cylinder, which consequently moves in relation to the tube. 

Assuming axisymmetry, we denote a cylindrical coordinate system $(x,r)$ with origins at the center of the opening of the tube.  We denote time $t$, liquid velocity $(u,v)$, liquid gauge pressure $p$, liquid viscosity $\mu $, liquid density $\rho $ and volume flux ${v_q(t)}$ entering or existing the tube through the inlet. We denote the length of the cylinder which penetrated into the tube ${l_p}$, Young's modulus of the tube $E$,  tube radial deformation ${d_{r}}$, tube inner radius ${r_i}$ and outer radius ${r_o}$, tube length $l$,  tube wall thickness $w={r_o}-{r_i}$, inner cylinder radius at rest ${r_p}$. The gap between the cylinder and the tube at rest is ${h_0}={r_i}-{r_p}$.  ${f_e}(t)$ denotes external force applied on the cylinder. $\tilde z$ is an auxiliary moving coordinate located at the tip of the penetrating cylinder, and related to the $x$ coordinate by $\tilde z = {l_p}(t)-x$.

Hereafter, normalized variables are denoted by uppercase letters and characteristic parameters are denoted by lowercase letters with asterisks (e.g. if $f$ is a dimensional variable, ${f^{\rm{*}}}$ is the characteristic value of $f$ and $F = f/{f^{\rm{*}}}$ is the corresponding normalized variable).

We define the normalized coordinates $X,R$ (starting at the tube inlet), the normalized moving coordinate $\tilde Z$ (starting at the penetrated side of the cylinder), and time $T$
\begin{equation}
X = \frac{x}{l},\quad R=\frac{r}{r_i},\quad \tilde Z = \frac{{\tilde z}}{{{l_p}}}=\frac{{{l_p}-x}}{{{l_p}}},\quad T=\frac{t}{t^*}.
\label{scaling_1}
\end{equation}
The normalized radial deformation $D_r$, pressure $P$, external force $F_e$, and penetrated length $L_p$ are
\begin{equation}
{D_r} = \frac{{{d_r}}}{{d_r^*}},
\quad P = \frac{p}{{p^ * }},\quad {F_e} = \frac{{{f_e}}}{{f_e^ * }},\quad {L_p} = \frac{{{l_p}}}{{l_p^ * }}
\label{scaling_2}
\end{equation}
where the relations between $d_{r}^*$, $p^*$, $l_p^*$,  $f_e^*$  and $t^*$ are derived for various limits in the following sections. The ratio of initial gap $h_0$ to characteristic radial deformations is denoted by
\begin{equation}
{\lambda _h} = \frac{{{h_0}}}{{d_r^ * }}.
\end{equation}
Small ratios required in applying the lubrication approximation are
\begin{equation}
\frac{h_0}{r_i} \ll 1,\quad\frac{{{d_{r}^*+h_0}}}{{{l_p^*}}} \ll 1,\quad  \frac{{{d_{r}^*+h_0}}}{{{l_p^*}}}\frac{\rho u^* (h_0+d_r^*)}{\mu}\ll1
\label{UP200}
\end{equation}
corresponding to slender configuration in both axial and radial coordinates, along with negligible inertial effects. The elastic shell model requires the small ratios,
\begin{equation}
\frac{w}{{{r_i}}} \ll 1,\quad \frac{{{d_{r}^ *} }}{{{r_i}}} \ll 1
\end{equation}
corresponding to thin wall thickness and small elastic deformations.

The ratio between the penetrated length of the cylinder and the length of the tube is defined by
\begin{equation}
{\varepsilon_l} = \frac{{{l_p^ *} }}{l}<1.
\end{equation}
 For the majority of this work, we will use the leading-order relation between the pressure inside the tube and its radial deformation \citep{timoshenko1959theory},
\begin{equation}
P(X,T)= \frac{{Ewd^*_r}}{{{p^* r_i^2}}}{D_r}(x,t),
\label{pdr}
\end{equation}
and thus $p^*=d_r^*Ew/r_i^2$, and $P=D_r$ for all cases hereafter.

In order to obtain the governing equations, we apply the lubrication approximation for flow in the gap region between the cylinder and the tube. Normalizing the lubrication equations according to (\ref{scaling_1})-(\ref{scaling_2}), neglecting $O(h_0/r_i)$ terms, and using standard procedures \citep{leal2007advanced}, yields the relevant Reynolds equation 
\begin{multline}
{\frac{{\partial {D_r}}}{{\partial T}}}   -\frac{t^*Ew h_0^3}{r_i^2 12\mu (l_p^*)^2}\frac{\partial }{\partial \tilde Z}\left[\frac{1}{L_p^2 }\frac{\partial P}{\partial \tilde Z}\left( 1+\frac{D_r}{\lambda_h}\right)^3 \right] +\lambda_h\frac{\partial }{\partial \tilde Z} \left[ \frac{\partial {L_p}}{\partial T} \frac{1}{2L_p}\left( 1 +\frac{D_r}{\lambda_h}\right) \right] =0.
\label{Def_equation}
\end{multline}
The Reynolds equation is supplemented by two integral constraints. The first is integral mass conservation, given in scaled form by
\begin{equation}
L_p -L_p(0) - \frac{2 d_r^* l}{ r_i l_p^*} \int_0^1 {D_r dX}  + \frac{{h_0^3{p^*}{t^*}}}{{6{r_p}\mu {l_p^*}^2}} \left(\int_0^T \frac{1}{L_p}\frac{\partial P}{\partial \tilde Z}\bigg|_{\tilde Z = 1}d\tilde T \right)  =0,
\label{Mass_eq}
\end{equation}
where gauge pressure at the inlet is zero, $P(X=0)={D_r}(X = 0) = 0$. The first term in (\ref{Mass_eq}) is the volume of liquid displaced by the advancing cylinder, the second term is the volume change due to solid deformation and the last term is the total volume of the liquid which exited the system through the annular inlet since $T=0$.

The second constraint is scaled integral momentum equation on the inner cylinder, given by 
\begin{equation}
F_e-\frac{\pi r_p^2 p^*}{f_e^*}P(\tilde Z=0)-\frac{2\pi r_p \mu {l_p^*}^2}{ d_r^* f_e^* t^*}\left[ {L_p}\frac{{\partial {L_p}}}{{\partial T}}\int_0^1\frac{d\tilde{Z}}{\lambda_h+D_r} \right]=0,
\label{Force_eq}
\end{equation}
where the first term is the external force, the second term is the pressure at the base of the penetrating cylinder ($\tilde Z=0$)  and the third term is shear stress of the cylinder wall ($R=R_p$).

Since a general analytic solution of the above equations is not available, approximated solutions for various limits will be pursued in the following sections in order to provide insight regarding the examined configuration. We begin by examining the simplest linearized limit.

\section{Linearized insertion \& extraction dynamics, ${|f_e|}\ll E h_0 w $} \label{SDFLH}
In this section we begin studying this problem by examining the simplified linearized case of small deformations where ${\lambda _h} \gg 1$. 
For $\lambda_h=h_0/d_r^*\gg1$ and  $h_0 l/r_i l_p^*\lesssim 1$, order of magnitude analysis of (\ref{Mass_eq}) yields
\begin{equation}
t^*=\frac{{6{r_p}\mu {l_p^*}^2}}{h_0^3{p^*}},
\label{tstar}
\end{equation}
and substituting (\ref{tstar}) into (\ref{Force_eq}) yields $p^*=f_e^*/\pi r_p^2$ and $P(\tilde Z=0)=F_e+O(h_0^2/r_p^2)$.

In order to obtain approximated solution of (\ref{Def_equation}), under the integral constraints (\ref{Mass_eq}) and (\ref{Force_eq}), we introduce the regular asymptotic expansions
\begin{subequations}
\begin{equation}
  P\left( {\tilde Z,T} \right)=D_r\left( {\tilde Z,T} \right) =D_{r,0}+\lambda_h^{-1}D_{r,1}
\end{equation}and
\begin{equation}
  {L_p}\left( T \right)=L_{p,0}+\lambda_h^{-1}L_{p,1}.
\end{equation}
\label{asymptotic_expan}
 \end{subequations}

Substituting the expansions (\ref{asymptotic_expan}) and collecting terms for each order, the leading order of (\ref{Def_equation})-(\ref{Force_eq}) is given by

\begin{equation}
\quad \frac{\partial^2 P_0}{\partial \tilde Z^2}=0,
\quad { {L_{p,0}}} - {L_p(0)} +\int_0^T \frac{1}{L_{p,0}}\frac{\partial P_0}{\partial \tilde Z}\bigg|_{\tilde Z = 1}d\tilde T  = 0,
\label{leadingO}
\end{equation}
along with the boundary conditions $P_0(\tilde Z=0,T)=F_e$ and $P_0(\tilde Z=1,T)= 0$. The leading order equations are readily solved, yielding
\begin{equation}
{{L_{p,0}}(T) = \sqrt {2 {\int\limits_0^T {{F_e}(\tilde{T})d\tilde{T}} }  + {L_p(0)}^2} },\quad {P_0}\left( {\tilde Z,T} \right) = \left( {1 - \tilde Z} \right){F_e}.
\label{(A)}
\end{equation}

Substituting the expansions (\ref{asymptotic_expan}) and collecting terms of order $O\left({\lambda_h}^{-1}\right)$ of (\ref{Def_equation})-(\ref{Force_eq}), yields the first-order equations
\begin{equation}
\frac{\partial ^2 P_1}{\partial \tilde{Z}^2} = 0,
\quad { {L_{p,1}}} - {\int_0^1 {{D_{r,0}}dX} }-  {\int \limits_0^T   \frac{{\partial {P_0}}}{{\partial \tilde Z}}\bigg|_{\tilde Z = 1}\frac{{{{ {{L_{p,1}}} }}}}{{{{ {{L_{p,0}^2}} }}}}d\tilde T}  = 0,
\label{M1}
\end{equation}
along with the homogeneous boundary conditions $P_1(\tilde{Z}=0,T)=0$ and $P_1(\tilde{Z}=1,T) = 0$.

In order to solve (\ref{M1}), the integral of the leading order deformation is split into two regions:
\begin{equation}
\int_0^1 {{D_r}dX} = {\varepsilon_l}{L_p}\int_0^1 {{D_r}d\tilde Z} + \int_{{\varepsilon_l}{L_p}}^1 {{D_r}dX},
\label{spldef}
\end{equation}
where the first region represents the penetrated region (see region $\tilde z >0$ in Figure 1) and the second region represents the unpenetrated region of the tube in which the pressure is approximately constant and equals $P_0(\tilde Z=0,T)=F_e$. Substituting (\ref{pdr}) into (\ref{spldef}), yields
\begin{equation} 
\begin{split}
\int_0^1 {{D_{r,0}}dX} = {\varepsilon_l}{ {L_{p,0}}}\int_0^1 {\left( {1 - \tilde Z} \right){F_e }d\tilde Z} + {F_e}\left( {1 - {\varepsilon_l}{{ {{L_{p,0}}} }}} \right)= F_e\left( {1 - \frac{{{\varepsilon_l}}}{2}{{ {{L_{p,0}}} }}} \right).
\label{Dr to P}
\end{split} 
\end{equation} 
Differentiating (\ref{M1}) with respect to $T$ after substituting the expression obtained in (\ref{Dr to P}), as well as the solution of $P_{0}(\tilde{Z},T)$ from (\ref{(A)}), yields the ODE governing ${{L_{p,1}}}$
\begin{equation} \begin{split}
\frac{{\partial {{ {{L_{p,1}}} }}}}{{\partial T}} + \frac{{ {{ {{F_{e}}} }}}}{{L_{p,0}^2}}{ {L_{p,1}}} = \frac{\partial}{\partial T}\left[F_e\left(1-\varepsilon_p\frac{L_{p,0}}{2}\right)\right] 
\label{B}
\end{split} \end{equation} 
and thus the first-order correction, which accounts for effects of elasticity, is
\begin{equation}
L_{p,1}=e^{-\int_{0}^{T}{\frac{F_e}{L_{p,0}^2}d\tau}}\int_{0}^{T}{\frac{\partial}{\partial T}\left[F_e\left(1-\varepsilon_p \frac{L_{p,0}}{2}\right)\right]e^{\int_{0}^{\tilde{T}}{\frac{F_e}{L_{p,0}^2}d\tau}}d\tilde{T}},\quad P_1\left({\tilde{Z},T} \right)= 0.
\label{UP300}
\end{equation}

\begin{figure} \centering
\includegraphics[width=\textwidth]{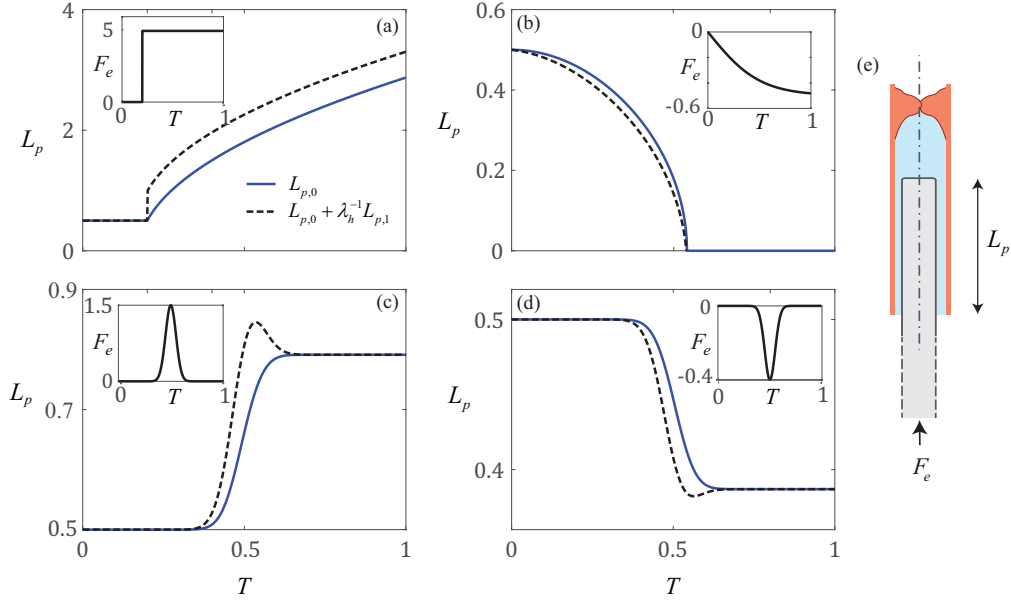}
\captionsetup{justification=justified}
\caption{\textbf{Motion due to various external force profiles.} Smooth lines denote (rigid) leading order solution (\ref{(A)}) and dashed lines denote first order solutions which include effects of elasticity (\ref{UP300}). In all panels $L_{p}(T=0)=0.5$, $\varepsilon_p=0.1$, $h_0/r_p=0.1$ and  $\lambda_h=10$. The examined force profiles (see inserts) are  ${F_e(T)} =5\cdot H(T-0.2)$ (a), $-0.5\cdot \tanh(2T)$ (b), $  1.5\cdot{\exp [{{{{{\left( {-T - 0.5 } \right)}^2}}}/{{2\cdot{0.05 ^2}}}}]}$ (c),  $ -0.4\cdot {\exp[{{{{{\left( {-T - 0.5 } \right)}^2}}}/{{2\cdot{0.05 ^2}}}}]}$ (d). The leading order motion is always is the direction of the external force. However, elastic effect yield extrema points for non-monotonous forces (panels c,d), along with a motion of the cylinder in a direction opposite to the external force.}
\label{PUSH_SMALL_ALL}
\end{figure}

We illustrate these results in Figure \ref{PUSH_SMALL_ALL}, showing  the motion of the cylinder for various external force profiles. The leading order solutions $L_{p,0}$ given by (\ref{(A)})  represent reference rigid configurations and are marked by solid lines. The effect of elasticity is presented by the difference between the leading and first-order solutions,  $L_{p,0}+\lambda_h^{-1}L_{p,1}$ given by (\ref{UP300}) and marked by dashed lines. The actuating external force vs. time profiles are presented as inserts within the panels. For all panels, normalized initial insertion of the cylinder is $L_{p}(T=0)=0.5$, ratio of cylinder to tube length is $\varepsilon_p=l_p^*/l=0.1$, initial gap to tube radius ratio is $h_0/r_p=0.1$ and initial gap to radial displacement ratio is  $\lambda_h=h_0/d_r^*=10$.

Panels (a) and (b) present the motion of the inner cylinder due to a sudden positive forcing (${F_e} =5 H(T-0.2)$)  and gradual negative forcing ($F_e=-0.5 \tanh(2T)$). For positive forcing, as expected, elasticity increases the penetration of the cylinder into the tube. While the positive pressure reduces viscous resistance via increased gap thickness, the dominant effect is related to deformation created ahead of the inner cylinder, which allows penetration of the cylinder due to displacement of liquid via the increased cross-sectional areas. This is evident also in panel (b), where elasticity decrease $L_p$, even even though negative pressure increases viscous resistance in this case. Panels (c) and (d) present forcing patterns of respectively positive and negative Gaussian profiles with mean of 0.5 and variance of $0.05^2$. In this case, tube elasticity creates a maxima point, and thus change the direction of motion of the inner cylinder as the elastic energy stored in the tube is gradually released.

The linearized limit used in this section allowed short preliminary examination of the system dynamics for a simplified case, showing the effect of elasticity on viscous resistance, as well as the effect of potential elastic energy. We proceed to examine non-linear configurations in the following sections.

\section{Nonlinear insertion dynamics, ${f_e}\gg E h_0 w $} \label{insertionPart}

\begin{figure}
\centering
\includegraphics[width=0.9\textwidth]{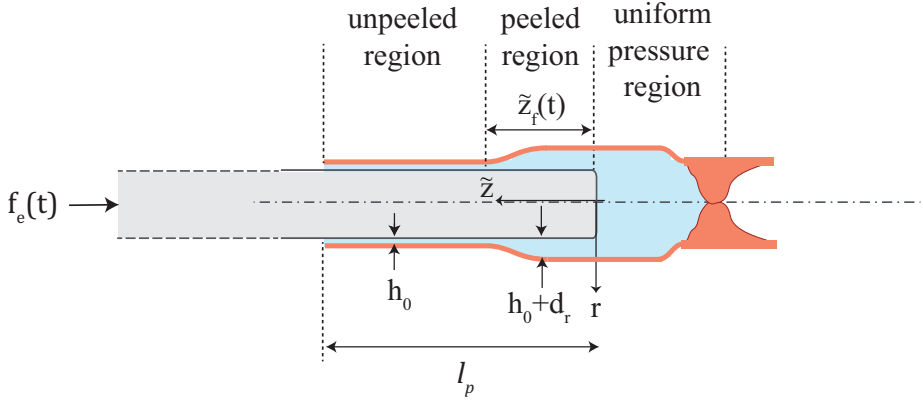}
\captionsetup{justification=justified}
\caption{\textbf{Illustration of the configuration for an external insertion force, in the limit of large deformation.} We distinguish between three domains: the unpeeled region, the peeled region and the uniformly pressurized region. These domains change in time due to the advance of the cylinder, $l_{p}(t)$, and the propagation of the peeling front, ${\tilde z_{F}}(t)$.}
\label{figure_compress}
\end{figure}

This section examines a sufficiently large positive external force which inserts the cylinder into the tube and creates large deformations compared with the initial gap between the inner cylinder and the external tube (i.e. $\lambda_h=d_r^*/h_0\gg1$). In this limit, as shown below, the Reynolds' evolution equation (\ref{Def_equation}) is reduced to a Porous-Medium-Equation, a nonlinear diffusion equation characterized by solutions with distinct non-smooth fronts. The non-smooth front separates between a pressurized deformed region and a region with trivial solution of zero gauge pressure and deformation in which the liquid pressure did not propagate yet.  We refer the reader to \cite{vazquez2007porous} for a detailed discussion of such equations. This section will focus on dynamics involving such a front, and thus an additional geometric division is required, separating the region before the front and after the front. This division is presented in figure \ref{figure_compress}.

The location of the front is denoted hereafter as ${\tilde z_F}$, and in normalized form by
\begin{equation}
{\tilde Z_F} = \frac{{{{\tilde z}_F}}} {{{l_p}}}.
\end{equation}
The three different domains are presented in Figure \ref{figure_compress} and include: (I.) The unpeeled region  ${{\tilde z}_F}(t) < \tilde z \le {l_p}(t)$ to which the deformation front has not yet reached.(II.) The peeled lubrication region  $0 < \tilde z \le {{\tilde z}_F}(t)$, and (III.) ${l_p}(t) < x \le l$, the uniformly pressurized bulk of the tube. 

%
\subsection{Scaling \& derivation of governing equations}
The above physical limit allows to simplify the governing Reynolds' equation, as well as the integral mass and momentum conservation constraints (\ref{Mass_eq})-(\ref{Force_eq}).  

The compact support of the fluidic front (see discussion below (\ref{Def_equation_nonlinear_OUTER_PME})) allows to simplify  integral mass conservation to
\begin{equation} \begin{split}
{L_p} - {L_p(0)} = \frac{2 r_i d_r^* l}{r_p^2 l_p^*} \left[(1-\varepsilon_l L_p) P(\tilde Z =0)+\varepsilon_l {{L_p}\int_0^{{{\tilde Z}_F}} {{D_{r}}d\tilde Z} } \right].
\label{3.57}
\end{split} \end{equation} 
For the integral force conservation equation, the effect of shear at the peeled region II can be neglected compared with the unpeeled region I, thus simplifying (\ref{Force_eq}) to 
\begin{equation}
{{F_e} - \frac{{\pi {r_p}^2{p^ *} }}{{{f_e^ *} }}{P(\tilde{Z}=0)} - \frac{{2\pi {r_p}\mu {l_p^*}^2}}{{{f_e^ *} {h_0}{t^*}}}\left[ {{L_p}\frac{{\partial {L_p}}}{{\partial T}}(1 - {{\tilde Z}_F})} \right] = 0}.
\label{F_e_gov}
\end{equation}

Together with (\ref{pdr}) and $p^*=f_e^*/\pi r_p^2$  we can simplify (\ref{3.57}) to
\begin{equation}
{{P(\tilde{Z}=0)}= \frac{{E{w}{{l_p^*}}}}{{2l{p^ * r_p} }}\left[ {\frac{{{L_p} - {L_p(0)}}}{{1 - {\varepsilon_l}{L_p}}}} \right] - {\varepsilon_l}\left[ {\frac{{{L_p}\int_0^{{{\tilde Z}_F}} {{D_{r}}d\tilde Z} }}{{1 - {\varepsilon_l}{L_p}}}} \right]},
\label{P_t_gov}
\end{equation}
and scaling of (\ref{P_t_gov}) therefore yields $l_p^*=p^ * 2l{ r_p}/Ew$. Substituting (\ref{P_t_gov}) into (\ref{F_e_gov}),  the  set of equations governing the large deformation limit are
\begin{subequations}
\begin{equation}
\begin{split}
{F_e} = \left[ {\frac{{{L_p}\left( {1 - {\varepsilon_l}\int_0^{{{\tilde Z}_F(T)}} {{D_{r}}d\tilde Z} } \right) - {L_p(0)}}}{{1 - {\varepsilon_l}{L_p}}}} \right] + \Pi_1\left[ {{L_p}\frac{{\partial {L_p}}}{{\partial T}}(1 - {{\tilde Z}_F(T)})} \right],
\label{ODE}
\end{split}
\end{equation}
and
\begin{equation}
{\frac{{\partial {D_r}}}{{\partial T}}}   =\Pi_2\frac{\partial }{\partial \tilde Z}\left(\frac{1}{L_p^2 }\frac{\partial P}{\partial \tilde Z} D_r^3 \right) -\frac{\partial }{\partial \tilde Z} \left( \frac{\partial {L_p}}{\partial T} \frac{D_r}{2L_p} \right),
\label{Def_equation_nonlinear}
\end{equation}
where, by using the obtained relations between the characteristic values ($f_e=p^* \pi  r_p^2 $, $l_p^* =p^*2lr_p/Ew$ and $d_r^*=p^*r_i^2/Ew$)  $\Pi_1$ and $\Pi_2$ can be presented in terms of $f_e^*$, $t^*$ and known geometric and physical parameters of the problem, 
\begin{equation}
\Pi_1=\frac{8l^2 \mu f_e^*}{\pi r_p h_0 (Ew)^2 t^* },\quad \Pi_2=\frac{t^* f_e^*r_i^4}{\pi 48\mu l^2r_p^4}.
\end{equation}
\label{GOV_NON_INSERT}
\end{subequations}

For a prescribed $f_e^*$, two different time-scales are evident in (\ref{GOV_NON_INSERT}). The first is 
\begin{equation}
{t_{\textrm{shear}}^ *} =\frac{ 8l^2 \mu f_e^*}{\pi r_p h_0 (Ew)^2 },
\end{equation}
which is the time-scale in which shear is a leading-order term in (\ref{ODE}), (shear is described by the second RHS term). For $t^*\ll {t_{\textrm{shear}}^ *}$, pressure effects can be neglected in (\ref{ODE}) and the dominant balance would be between the external force and viscous shear.

The second time-scale is 
\begin{equation}
{t_{\textrm{peeling}}^ *} = \frac{\pi 48\ \mu l^2 r_p^4}{f_e^* r_i^4 },
\label{t_peeling}
\end{equation}
and represents the time-scale in which pressure-driven flow is a dominant term in region II (see first RHS term in (\ref{Def_equation_nonlinear})). 

\subsection{Matched asymptotics}
The value of $\Pi_1$ decreases with $t^*$ while $\Pi_2$ increases with $t^*$, thus indicating that early time-dynamics are governed by shear, while late-time dynamics are governed by pressurization of the liquid within the tube. Hereafter, in order to proceed, we limit our focus to configurations with the small ratios of
\begin{equation}\label{ass2}
 \varepsilon_t=\frac{t_{\textrm{shear}}^ *}{{t_{\textrm{peeling}}^ *}}\ll1,\quad   \varepsilon_l=\frac{l_p^*}{l}\ll1.
\end{equation} 
Additionally, this section examines the response of the configuration to a specific case of suddenly applied external force of the form
\begin{equation}
F_e=H(T),
\end{equation}
where $H(T)$ is the heaviside function.

\subsubsection{Outer region}
For time-scales of $t^*=t^*_{\textrm{peeling}}$, we obtain $\Pi_2=1$ and $\Pi_1=\varepsilon_t$ and thus the leading order (\ref{GOV_NON_INSERT}) is simplified to
\begin{subequations}
\begin{equation}
\begin{split}
{F_e} = \left[ {\frac{{{L_p}\left( {1 - {\varepsilon_l}\int_0^{{{\tilde Z}_F(T)}} {{D_{r}}d\tilde Z} } \right) - {L_p(0)}}}{{1 - {\varepsilon_l}{L_p}}}} \right] + O(\varepsilon_t),
\label{ODE_OUTER}
\end{split}
\end{equation}
and
\begin{equation}
{\frac{{\partial {D_r}}}{{\partial T}}}   =\frac{\partial }{\partial \tilde Z}\left(\frac{1}{L_p^2 }\frac{\partial P}{\partial \tilde Z} D_r^3 \right) -\frac{\partial }{\partial \tilde Z} \left( \frac{\partial {L_p}}{\partial T} \frac{D_r}{2L_p} \right).
\label{Def_equation_nonlinear_OUTER}
\end{equation}
\label{GOV_NON_INSERT_OUTER}
\end{subequations}
Within the outer-region, we apply regular asymptotics with regard to the second small parameter $\varepsilon_l$ (which is $\gg \varepsilon_t$). The asymptotic expansions for both $L_p$ and $P=D_r$ are defined by
\begin{equation}
L_p=L_{p,0}+\varepsilon_l L_{p,1},\quad P=D_r=D_{r,0}+\varepsilon_l D_{r,1}.
\end{equation}
Applying standard asymptotic procedure on (\ref{ODE_OUTER}) yields 
\begin{equation}
L_p=F_e+L_p(0)+\varepsilon_l (F_e+L_p(0))\left( \int_0^{\tilde Z_{F(T)}}{D_{r,0}d\tilde Z} -F_e\right)
\label{asym_LP}
\end{equation}
which depends only on the leading-order deformation $D_{r,0}$. Substituting $L_p$ from (\ref{asym_LP}) into (\ref{Def_equation_nonlinear_OUTER}),  yields the PDE governing $D_{r,0}$,
\begin{equation}
\frac{\partial D_{r,0}}{\partial T}   =\frac{1}{4(F_e+L_p(0))^2 }\frac{\partial ^2}{\partial \tilde Z^2}\left( D_{r,0}^4 \right),
\label{Def_equation_nonlinear_OUTER_PME}
\end{equation}which is a Porous-Medium-Equation of order 4, supplemented by the initial condition of $D_{r,0}(\tilde Z,0)=0$ and boundary condition of $P_0=D_{r,0}(0,T)=1$ (we set $F_e=H(T>0)=1$).

Self-similar treatment of the above equation (\ref{Def_equation_nonlinear_OUTER_PME}) is possible, following the approach presented by \cite{zel1950towards} and \cite{barenblatt1952some}. Defining the self-similar variable 
\begin{equation}
\eta=2(F_e+L_p(0))\frac{Z}{T^{1/2}}
\label{eta_def}
\end{equation}
yields an ODE for $f(\eta)=D_{r,0}$ 
\begin{subequations}
\begin{equation}
(f^4(\eta))''+\frac{1}{2}\eta f'(\eta)=0,
\end{equation}
along with initial and boundary conditions

\begin{equation}
\quad f(0)=1,\quad f(\eta_f)=0.
\end{equation}
\label{selfsim}
\end{subequations}
Equation (\ref{selfsim}) can be solved numerically \citep[see][]{vazquez2007porous,elbaz2016axial}, yielding  
\begin{equation}
\eta _f=1.704,\quad \int_0^{\eta_f} {f(\eta)d\eta}=1.305.
\label{numvals}
\end{equation}
Substituting (\ref{eta_def}) into (\ref{numvals}), we obtain the front location
\begin{equation}
Z_f=0.852\frac{T^{1/2}}{F_e+L_p(0)}
\label{front}
\end{equation}
as well as the mass-flux entering region II 
\begin{equation}
 \int_0^{\tilde Z_f} {D_{r,0}d\tilde Z}=\frac{\partial \tilde Z}{\partial \eta} \int_0^{\eta_f} {f(\eta)d\eta} =0.652\frac{T^{1/2}}{F_e+L_p(0)}.
 \label{mass_ZT}
\end{equation}
Substituting (\ref{mass_ZT}) into (\ref{asym_LP}), yields the solution for $L_p$ for the outer-region
\begin{equation}
L_p=F_e+L_p(0)+\varepsilon_l \left[ 0.652(T^{1/2}) -F_e(F_e+L_p(0))\right].
\end{equation}
The initial condition $L_p(0)$ of the outer region needs to be related, via matching, to the inner-region solution derived in the next subsection. 

\subsubsection{Inner region}
We introduce the rescaled inner region coordinate
\begin{equation}
\bar T=\frac{T}{\varepsilon_t}.
\end{equation}
For inner region time-scale $t^*\sim \varepsilon_t  t^*_{\textrm{peeling}}$,  we can estimate the location of the peeling front is $\tilde Z_F(\varepsilon_t \bar T)\sim O(\varepsilon_t)$. Thus, the leading-order $O(1)$ balance in (\ref{ODE}) yields the inner-region solution $\bar L_p$ 
\begin{equation}
 \bar L_p\frac{\partial {\bar L_p}}{\partial \bar  T} = - \bar L_p +\bar L_p(0)+F_e  ,
\end{equation}
which is the Abel equation of the second kind \citep{zaitsev2002handbook}, for which a closed-form solution for Heaviside-function force, ${F_e}(T) =H(T)$ is available. The inner-region dynamics are thus given by
\begin{equation}
{L_p}(\bar T) = (L_p(0)+F_e)\left( {1 + {\rm{W}}\left[ -{  \frac{1}{L_p(0)+F_e}{\exp{\left(-1-\frac{\bar T+C}{L_p(0)+F_e} \right)}}} \right]} \right),
\end{equation}
where $W(T)$ is the Lambert-$W$ function \citep{weisstein2002lambert}. The initial condition ${L_p}\left( {T = 0} \right) = {L_p(0)}$ gives the constant $C$
\begin{equation}
C = -(F_e+L_p(0))\rm{Ln}(F_e)-L_p(0).
\end{equation}
\subsubsection{Uniform solution}
Matching between the inner and outer regions is required in order to obtain a uniform asymptotic solution. This yields the requirement 
\begin{equation}
\lim_{\bar T\rightarrow\infty} \bar L_p = \lim_{ T\rightarrow0}  L_p 
\end{equation}
and thus $\bar L_p(0)+F_e=L_p(0)+F_e$, and $\bar L_p(0)=L_p(0)$. The composite expansion is therefore given by $\bar L_p (\bar T)+L_p (T)-(L_p(0)+F_e)$, yielding
\begin{multline}
L_{p,\rm{uniform}}=(L_p(0)+F_e)\Bigg\{ 1+\varepsilon_l \left[ \frac{0.652(T^{1/2})}{F_e+L_p(0)} -F_e\right]\\ + {\rm{W}}\left[ -{  \frac{1}{L_p(0)+F_e}{\exp{\left(-1-\frac{ T/\varepsilon_t+C}{L_p(0)+F_e} \right)}}} \right] \Bigg\},
\label{unifromLP}
\end{multline}
which is the uniform solution representing the response of the configuration to a sudden external load. The above solution incorporates the effect of a propagating front. However, the location of the front will reach the end of the tube for $T>[(F_e+L_p(0))/0.852]^2$ (see equation (\ref{front})). Before proceeding to discuss and present the uniform solution (\ref{unifromLP}), we will approximate the dynamics of the post-peeling regime and connected it to the uniform solution presented above.

\subsubsection{Post-peeling dynamics}
After the peeling front reached the outlet, the entire tube is peeled and the viscous resistance can be approximated by the quasi-steady deformation solution of the Reynolds' equation (\ref{Def_equation_nonlinear_OUTER_PME})
\begin{equation}
    D_r\approx(1-\tilde Z)^{1/4} P(0)
\label{postP}
\end{equation}
for $t^*\sim t^*_{\rm{peeling}}$, integral force balance yields $P(0)=F_e$ and integral conservation of mass yields

\begin{equation}
\frac{\partial L_p}{\partial T}=\varepsilon_l\frac{F_e^4}{L_p(t) }
\label{postLP}
\end{equation}
which is obtained by substituting (\ref{postP}) into the second RHS term of (\ref{Mass_eq}) and deriving with regards to time. Integrating (\ref{postLP}), we can obtain the post-peeling solution
\begin{equation}
L_p=\sqrt{L_p^2(T=T_B)+2\varepsilon_lT F_e^4}
\label{postpeeling}
\end{equation}
valid for $T>T_B=[(F_e+L_p(0))/0.852]^2$. 

\subsection{Summary of nonlinear insertion dynamics}
\begin{figure} \centering
\includegraphics[width=1\textwidth]{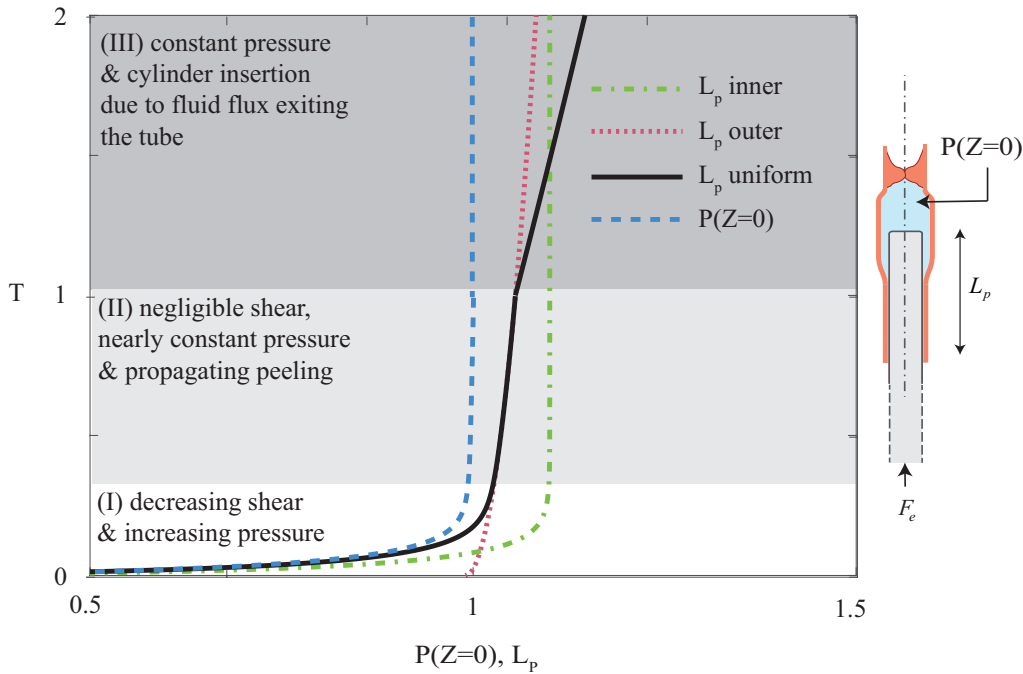}
\captionsetup{justification=justified}
\caption{\textbf{The location of the cylinder within the tube $L_p$, and liquid pressure ahead of the cylinder $P(Z=0)$, vs. time $T$.} The presented results are for configuration defined by $L_p(T=0)=0.1$, $\varepsilon_l=0.1$, $\varepsilon_t=0.05$ and $F_e=H(T)$. The white, light grey and dark grey parts denote the inner, outer and post-peeling regions, respectively.}
\label{UNI}
\end{figure}

We illustrate these results in Figure \ref{UNI}, which presents the location of the cylinder within the tube ($L_p$) vs. time, for the normalized parameters $F_e=H(T)$, $\varepsilon_l=0.1$, $\varepsilon_t=0.05$, and initial condition $L_p(0)=0.1$. The figure presents the uniform solution (\ref{unifromLP}) for $0<T<T_b$ and the post-peeling solution (\ref{postpeeling}) for $T>T_B$, where $T_b=[(F_e+L_p(0))/0.852]^2$. The response of the examined configuration to a suddenly applied insertion force, at the large deformation limit ($\lambda_h=d_r^*/h_0\gg1$), is shown to involve three distinct regimes. In the first regime (inner-regime, \S4.2.2) the motion of the cylinder is governed by balance between shear stress, fluidic pressure and the external force, while the effect of viscous peeling is negligible. As $t$ increases, the effect of shear decreases while the fluidic pressure increases. In the second regime (outer-regime, \S4.2.1) the cylinder decelerates and the external force is balanced by fluidic pressure ahead of the cylinder, while effects of shear are negligible. The motion of the cylinder in this regime is governed by balance between the fluidic volume displaced by insertion of the cylinder and the fluidic volume required to peel the inner cylinder from the external tube. Finally, as the peeling front reaches the inlet of the tube, a third post-peeling regime is obtained (\S4.2.4). In this regime the cylinder motion is governed by balance between  balance between the volume displaced by the cylinder and the  fluidic flow outside of the configuration.

The next section we proceed to examine the opposite case of extraction of the cylinder from the tube.

\section{Non-linear extraction dynamics, $f_e\lessapprox - E w h_0$}

This section will examine the forced extraction of an inner cylinder from a liquid-filled tube. In this case a negative external force ${f_e}(t)<0$ creates a negative gauge pressure within the tube, and thus negative deformations. We will focus on the non-linear limit involving negative deformations of the order of $h_0$ or greater (where $h_0$ is the initial gap between the cylinder and the tube).

\begin{figure} \centering
\includegraphics[width=0.7\textwidth]{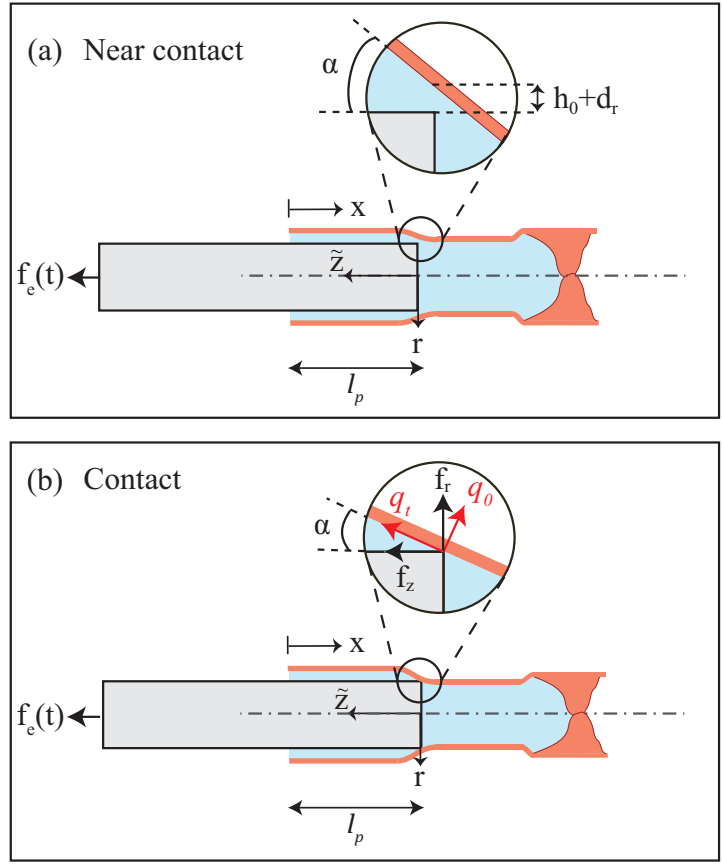}
\captionsetup{justification=justified}
\caption{\textbf{Illustration of the configuration at extraction for (a) near contact and (b) contact cases.} Panel (a) presents the near contact case, where the gap at $h_0+d_r(\tilde z=0)\rightarrow 0^+$, thus nearly separating the liquid into two regions. The slope at $\tilde z=0$ is given by $\alpha$.  Panel (b) presents contact, where the gap $h_0+d_r(\tilde z=0)=0$, and additional friction forces $(f_r,f_z)$ act on the elastic tube. The forces ($q_t,q_0$) are projections of the friction into direction normal and tangent  to the surface of  the elastic tube.
\label{PullPD}}
\end{figure}

The  dynamics during extraction can be described by the previously derived results for insertion for the small deformation limit ($d^*_r / h_0 \ll 1$, see \S3), as well as for the initial shear-dominated inner-region large deformation limit ($d^*_r / h_0 \gg 1$, $t^*\sim 8l^2\mu f_e^*/\pi r_p h_0 E^2 w^2$ see \S4.2.2). However, outer-region solutions at the large deformation limit ($d^*_r/h_0\gg1$, $t^*\gg 8l^2\mu f_e^*/\pi r_p h_0 E^2 w^2$) exhibit essentially different dynamics. The negative gauge pressure created during extraction closes the gap between the cylinder and the tube, and may create contact between the two solids. This section will examine the case of nearly contacting cylinder and tube (see Fig \ref{PullPD}a, \S5.1) and the case of contact (see Fig \ref{PullPD}b, \S 5.2). 


\subsection{Near contact, $f_e\rightarrow (-2\pi Ew h_0r_p^2/r_i^2)^+$}
For the limit of near contact between the tube and the cylinder (see Fig. \ref{PullPD}a) we define the normalized gap at  $X=\varepsilon_l L_p$  by 
\begin{equation}
D_r|_{X=\varepsilon L_p}+1=\varepsilon_{NC}H_{NC},\quad \varepsilon_{NC}=\ll1
\end{equation}
where $H_{NC}$ is the scaled minimal gap and the small parameter $\varepsilon_{NC}$ will be later related to the extraction force (alternatively, $H_{NC}=(d_r|_{x=l_p} +h_0)/(\varepsilon_{NC}h_0)\sim O(1)$). This limit can be leverage to the simplification of the governing equations, allowing analytical treatment. While the governing integral conservation equations of mass and momentum are largely unchanged, the equations governing the flow field and deformation are significantly modified.

\subsubsection{Governing equations}
The rapid change in fluid pressure near $X=\varepsilon_l L_p$  for $\varepsilon_{NC}\rightarrow 0$ requires to include elastic bending effects in the $X$ direction. Following \cite{timoshenko1959theory}, the deformation of a circular cylindrical tube loaded axisymmetrically is described by 
\begin{equation}
K\frac{{{\partial ^4}{D_r}\left( {X,T} \right)}}{{\partial {X^4}}} + {D_r}\left( {X,T} \right)\sim P\left( {X,T} \right),
\label{Nor_tim}
\end{equation}
where 
\begin{equation}
K \equiv \frac{{w^2 r_i^2}}{{12l^4\left( {1 - {\nu ^2}} \right)}},
\end{equation}
the deformation $D$ is scaled by $d_r^*=h_0$ and the pressure $P$ is scaled by $p^*=Ewh_0/r_i^2$. The external force $F_e$ is thus scaled by $f^*=\pi r_p^2 p^*$. Equation (\ref{Nor_tim}) replaces the previous relation (\ref{pdr}) between fluid pressure and elastic deformation.

The limit of near contact is singular with regard to fluid resistance, and thus viscous resistance is defined by the small finite parameter $\varepsilon_{NC}$. Applying a Taylor series around  $X=\varepsilon_lL_p$ allows to simplify the pressure-flux relation to
\begin{multline}
  \dot Q= -\frac{\pi r_i p^* h_0^3}{3\mu\dot q^* l}\frac{\partial P}{\partial X} \left(D_r+1 \right)^3 \\ \approx -\frac{\pi r_i p^* h_0^3}{3\mu\dot q^* l}\frac{\partial P}{\partial X} \left(D_r|_{X=\varepsilon_l L_p}+1 +\frac{\partial D_r}{\partial X}\Big|_{X=\varepsilon L_p}(X-\varepsilon_l L_p)\right)^3 
\end{multline}
where $\dot Q$ is scaled by $\dot q^*=\pi r_i p^* h_0^3/3\mu\dot q^* l$.  Integrating from $X=\varepsilon L_p$ to $X$,  and extracting the pressure difference in term of flux $\dot Q$, yields
\begin{multline}
 P|_{X=\varepsilon_l L_p}-P|_{X}= \left(D_r|_{X=\varepsilon_l L_p}+1\right)^{-2}\left(-2\frac{\partial D_r}{\partial X}\right)^{-1} \dot Q -\\ \left(D_r|_{X=\varepsilon_l L_p}+1 +\frac{\partial D_r}{\partial X}\Big|_{X=\varepsilon_l L_p}(X-\varepsilon_l L_p)\right)^{-2}\left(-2\frac{\partial D_r}{\partial X}\right)^{-1} \dot Q  .
\end{multline}
For $\varepsilon_{NC}\rightarrow0$ the pressure difference $P|_{X=\varepsilon_l L_p}-P|_{X}$ asymptotes to a  constant, defined only by the conditions near $X=\varepsilon_l L_p$. Applying $P(X=0)=0$, and $D_r|_{X=\varepsilon_l L_p}+1=H_{NC}\varepsilon_{NC}\ll1$, the following relation is obtained
\begin{equation}
\dot Q \approx-2\varepsilon_{NC}^2\left(\frac{\partial D_r}{\partial X} H_{NC}P\right)\Bigg|_{X=\varepsilon_l L_p} ,
\end{equation}
representing the flux only by the conditions at $X=\varepsilon_l L_p$ (where $P(X=0)=0$).  Thus, the governing integral mass and momentum conservation equations for extraction are 
\begin{equation} 
{L_p} - {L_p(0)} =\frac{2 r_i h_0 l}{r_p^2 l_p^*}\int_0^1D_rdX-\frac{h_0^4 E w  t^*\varepsilon_{NC}^2}{3\mu  l_p^* r_p^2 l r_i}  \int\limits_0^T \left(\frac{\partial D_r}{\partial X} H_{NC}P\right)\Bigg|_{X=\varepsilon_l L_p} {d\tilde T }
,\label{ext_vol2}
\end{equation}
and
\begin{equation}
{{F_e} - {P|_{X=\varepsilon_l L_p}} - \frac{2\mu (l_p^*)^2 r_i^2}{r_p Ewh_0^2t^*}\left[ {{L_p}\frac{{\partial {L_p}}}{{\partial T}}\int_0^1\frac{d\tilde Z}{1+D_r}} \right] = 0}.
\label{F_e_gov_22}
\end{equation}
These equations are similar to \S4.1, with the exception of the last RHS in (\ref{ext_vol2}), representing mass flux $\dot Q$. Scaling of (\ref{ext_vol2}) and (\ref{F_e_gov_22}) yields
\begin{equation}
l_p^*=\frac{2 r_i h_0 l}{r_p^2 },
\end{equation}
and two dimensionless ratios 
\begin{equation}
\tilde \Pi_1=\frac{h_0^3 E w  \varepsilon_{NC}^2  t^*}{6\mu    l^2 r_i^2},\quad\tilde \Pi_2=\frac{8\mu l^2 r_i^4}{r_p^5  Ewt^*}. 
\end{equation}
Similarly to the case of nonlinear insertion dynamics, the ratio $\tilde \Pi_1$ decreases with $t^*$ while $\tilde \Pi_1$ increases with $t^*$, suggesting different early-time and late-time dynamics. Thus two corresponding time-scales are evident. The first is the time-scale of the initial shear-dominated regime
\begin{equation}
t^*_{\textrm{shear}}=\frac{8\mu l^2 r_i^4}{r_p^5  Ew}.
\end{equation}
The second time-scale is of the late-time motion due to flow through near-contact gap between the cylinder and the tube,
\begin{equation}
t^*_{\textrm{NC}}=\frac{6\mu l^2 r_i^2 }{Ew  h_0^3 \varepsilon_{NC}^2}.\label{tshear}
\end{equation}
For the examined configuration, the ratio between the two time-scales is a geometrically small parameter, given by
\begin{equation}
\varepsilon_t=\frac{t^*_{\textrm{shear}}}{t^*_{\textrm{NC}}}=\frac{8 h_0^3 r_i^2 \varepsilon_{NC}^2}{6r_p^5}\ll1
\end{equation}
and so $\varepsilon_t \ll \varepsilon_{NC}^2$ . 

Substituting $t^*=t^*_{NC}$, yields the governing equations
\begin{equation} 
L_p-L_p(0) =(1-\varepsilon_l L_p)  P|_{X=\varepsilon_l L_p} - \int\limits_0^T \left(\frac{\partial D_r}{\partial X} H_{NC}P\right)\Bigg|_{X=\varepsilon_l L_p} {d\tilde T }
,\label{ext_vol}
\end{equation}
and
\begin{equation}
F_e - P|_{X=\varepsilon_l L_p} -\varepsilon_t\left( L_p\frac{\partial L_p}{\partial T}\int_0^1\frac{d\tilde Z}{1+D_r}\right)= 0.
\label{F_e_gov_2}
\end{equation}
We proceed by asymptotic expansions, based on perturbations from the exact contact state $\varepsilon_{NC}=0$. We thus define the external extraction force by 
\begin{equation}
{F_e} =F_{e,0}(1- {\varepsilon _{NC}}),
\end{equation}
where $F_{e,0}=F_{e,contact}$ is the external extraction force creating exact contact $D_r|_{X=\varepsilon_l L_p}=-1$, and thus zero flux $\dot Q=0$.  In addition, we define the expansions for the fluidic pressure
\begin{equation}
{P} = {P_{0}} + {\varepsilon _{NC}}{P_{1}}.
\end{equation}
\label{LLL}

We apply a matched asymptotic scheme, and begin by solving for the outer-region $t\sim t^*_{NC}$  .

\subsubsection{Outer-region }
The leading-order solution is the exact contact case, where $F_e=F_{e,contact}$ , and after the initial transition regime (\S 4.2.2), the contact point between the solids separate the fluid into two domains, and thus $\dot Q=0$. The outer-region force balance (\ref{Force_eq}) is simplified to
\begin{equation}
 P_0|_{X=\varepsilon_l L_p}=F_{e,0},\quad  P_1|_{X=\varepsilon_l L_p}=-F_{e,0}
\label{F.B.0}
\end{equation}
and the pressure-field is 
\begin{equation}
P_0(X,T)+\varepsilon_{NC} P_1(X,T)= H(X-\varepsilon L_p) \left( F_{e,0}+\varepsilon_{NC} F_{e,1}\right),
\label{HeaviP0}
\end{equation}
where $H$ is the Heaviside function.  The solid deformation is governed by equation (\ref{Nor_tim}), along with the boundary conditions, ${{{\partial ^2}{D_r}}}{{/\partial {X^2}}} | _{X=0} =  {{{\partial ^3}{D_r}}}{{/\partial {X^3}}}| _{X=0} = {D_r}| _{X=1} = {{\partial {D_r}}}{{/\partial X}}| _{X=1} = 0.$ Deformation patterns may be obtained numerically and analytically. However, since the singularity for $\varepsilon_{NC}\rightarrow 0$ dictates that flux is governed by the conditions near $X=\varepsilon _l L_p$, only expressions for $D_{r}|_{X=\varepsilon_l L_p}$ and $\partial D_{r}/\partial X|_{X=\varepsilon_l L_p}$ are required. Due to linearity and symmetry considerations, the gap slope at $X=\varepsilon_l L_p$ can be simplified to
\begin{equation}
D_{r}|_{X=\varepsilon_l L_p}=\frac{1}{2}\left(P|_{X<\varepsilon_l L_p}+P|_{X>\varepsilon_l L_p})\right),\label{gap}
\end{equation}
and
\begin{equation}
\frac{\partial D_r}{\partial X}\Bigg|_{X=\varepsilon_l L_p}=\frac{\left(P|_{X>\varepsilon_l L_p}-P|_{X<\varepsilon_l L_p})\right)}{(64K)^{1/4}}.\label{drdx}
\end{equation}
Thus, the requirement of $D_{r,0}|_{X=\varepsilon_l L_p}=-1$ yields
\begin{equation}
F_{e,0}=P_{0}(\tilde{Z}=0)=-2.\label{pres}
\end{equation}
and $H_{NC}$ is
\begin{equation}
H_{NC}=1
\end{equation}
and
\begin{equation}
\frac{\partial D_{r,0}}{\partial X}\Bigg|_{X=\varepsilon_l L_p}+\varepsilon_{NC} \frac{\partial D_{r,1}}{\partial X}\Bigg|_{X=\varepsilon_l L_p}=\frac{ {F_{e,0}}(1-\varepsilon _{NC})}{64K^{1/4}}.
\end{equation}

Substituting (\ref{gap})-(\ref{pres}) into  (\ref{ext_vol}), $L_p$ to order $\varepsilon_{NC}^2$  is thus
\begin{equation} 
L_p =L_p(0)+ F_e -\frac{ F_e^2}{(64K)^{1/4}}T
.\label{ext_vol}
\end{equation}





\subsubsection{Matched solution}
The inner solution is identical to the insertion case, and thus is given in \S4.2.2. Matching the inner and outer solutions for near-contact extraction yields the requirement 
\begin{equation}
\lim_{\bar T\rightarrow\infty} \bar L_p = \lim_{ T\rightarrow0}  L_p 
\end{equation}
and thus $\bar L_p(0)+F_e=L_p(0)+F_e$, and $\bar L_p(0)=L_p(0)$. The composite expansion is therefore given by $\bar L_p (\bar T)+L_p (T)-(L_p(0)+F_e)$, and the uniform solution is
\begin{multline}
L_{p,\rm{uniform}}=(L_p(0)+F_e)\Bigg\{1- \frac{1}{L_p(0)+F_e}\frac{ F_e^2}{(64K)^{1/4}}T\\ + {\rm{W}}\left[ -{  \frac{1}{L_p(0)+F_e}{\exp{\left(-1-\frac{ T/\varepsilon_t-(F_e+L_p(0))\rm{Ln}(F_e)-L_p(0)}{L_p(0)+F_e} \right)}}} \right] \Bigg\},
\label{unifromLP}
\end{multline}
which is presents the response of the configuration to a sudden external load.

The uniform solution is plotted in figure 6 for several configurations. In all cases, the parameters of $\varepsilon_t=0.1$, $K=0.3$ and $F_e=-1.9$ (corresponding to $\varepsilon_{NC}=0.05$) are used. Panel 6(a) presents the motion of the cylinder for the initial conditions $L_p(T=0)=5$ (smooth line), along with the inner solution (dashed line) and outer solution (dashed-dotted line). After the initial inner dynamics, the singular effect of the near-contact dominants the motion and creates a steady extraction speed independent of $L_p$. Panel (b) presents various initial values of $L_p(T=0)$, showing that the inner-regions changes with $L_p(T=0)$, but not the extraction speed at the outer region. For the case of $L_p(T=0)=2\approx -F_e$, we see that total extraction from the tube occurs before reaching the outer-region. Thus, the rapid extraction eliminates deceasing the pressure sufficiently to create near-contact dynamics.

\begin{figure} \centering
\includegraphics[width=0.9\textwidth]{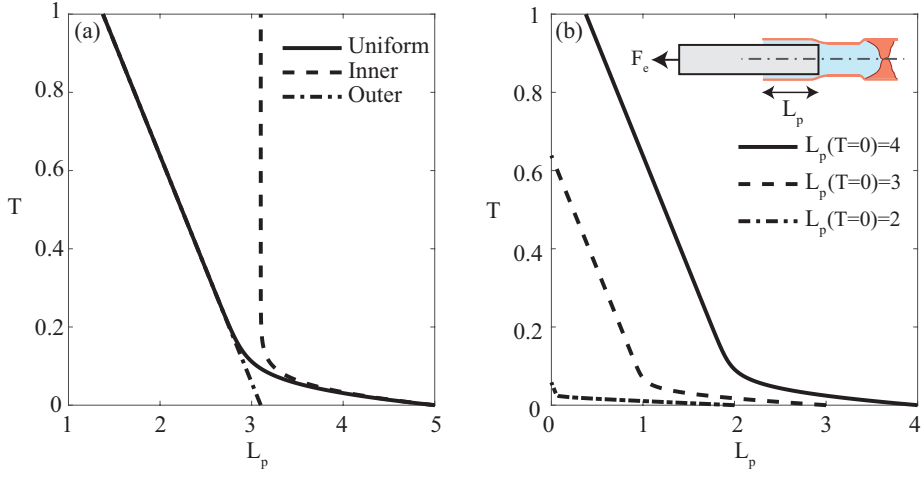}
\captionsetup{justification=justified}
\caption{\textbf{The location of the cylinder within the tube $L_p$ vs. time $T$ for the near contact limit.} All configurations are defined by $\varepsilon_t=0.1$, $K=0.3$ and $F_e=-1.9$ (corresponding to $\varepsilon_{NC}=0.05$). In panel (a) $L_p(T=0)=5$ and the different elements of the composite expansion are presented. Panel (b) presents various initial values of $L_p(T=0)$.}
\label{UNI_PULL}
\end{figure}

\subsection{Contact $f_e <- 2\pi Ew h_0r_p^2/r_i^2$}

Equation (\ref{gap}) indicates that for $F_e<-2$ (or in dimensionless form $f_e < -2\pi Ew h_0r_p^2/r_i^2$), there will be contact between the inner cylinder and the outer elastic tube.  In such a case, additional forces will be applied at $\tilde Z=0$, as illustrated in Fig. \ref{PullPD}(b), which will modify the system dynamics. The contact prevents fluid from exiting the tube, and applies normal and tangential forces on the elastic tube. Thus, the outer-region governing equations are simply 
\begin{equation}
L_p-L_p (0)=\int_{1-\varepsilon_l L_p}^{1}{D_r dX}
\end{equation}
and 
\begin{equation}
F_e-P|_{X=\varepsilon_l L_p}+F_z=0
\end{equation}
where $F_z=f_z/f^*$ is the additional force due to drag in the $\tilde Z$ direction, and using the scaling used in \S 5.1 ($d_r^*=h_0$, $l_p^*={2 r_i h_0 l}/{r_p^2 }$, $p^*=Ewh_0/r_i^2$ and $f_e^*=\pi r_p^2 p^*$). 

In dimensional terms, the friction force $f_z=q_t  \cos ⁡\alpha-q_0  \sin\alpha$, where $\tan\alpha=-(\partial d_r/\partial x)|_{x=l_p}$.  Since $\alpha\ll1$, we can approximate $\alpha\approx -(\partial d_r/\partial x)|_{x=l_p}$ and obtain
\begin{equation}
f_z=2\pi  r_i q_0 \left(\mu_f+\frac{\partial d_r}{\partial x}\bigg|_{x=l_p} \right).\label{fz}
\end{equation}
Calculation of the normal force acting on the elastic tube, $q_0$, is obtained from the requirement of $d_r|_{x=l_p}=-h_0$, where the deformation $d_r|_{x=l_p}$ is given by
\begin{equation}
d_r|_{x=l_p}=-q_0\frac{ r_i^2}{Ew }\left(\frac{{3\left( {1 - {\nu ^2}} \right)}}{{16 w^2 r_i^2}}\right)^{1/4}+p|_{x=l_p} \frac{r_i^2}{2Ew}\label{q0}
\end{equation}
representing the linear summation of the deformation due to a a localized normal force $q_0$ \citep[first RHS term, ][]{timoshenko1959theory} and the deformation due to liquid pressure (second RHS term, see equation (\ref{gap})).

We now calculate the normal force $q_0$ by requiring $d_r|_{x=l_p}=-h_0$ in equation (\ref{q0}). The obtained  $q_0$ is substituted into (\ref{fz}), along with ${\partial d_r}{\partial x}|_{x=l_p}$  calculated from (\ref{drdx}) (the localized force creates a symmetrical deformation, and does not affect the slope). This procedure yields the dimensional friction force $f_z$
\begin{equation}
f_z≈-\frac{2\pi Ew}{r_i}\left(h_0+p|_{x=l_p}\frac{r_i^2}{2Ew}\right)\left(\mu_f \left(\frac{{64w^2 r_i^2}}{{12\left( {1 - {\nu ^2}} \right)}}\right)^{1/4}+  p|_{x=l_p}\frac{r_i^2} { Ew} \right),
\end{equation}
or in normalized form, $F_z$
\begin{equation}
F_z=-2\left( 1+\frac{P|_{X=\varepsilon_lL_p}}{2}\right)\left(\hat \Pi_1 +\hat\Pi_2P|_{X=\varepsilon_lL_p}\right)
\end{equation}
where
\begin{equation}
\hat \Pi_1=\frac{\mu_f }{r_p^2}\left(\frac{{16 w^2 r_i^6}}{{3\left( {1 - {\nu ^2}} \right)}}\right)^{1/4}
,\quad \hat \Pi_2=\frac{r_i h_0}{r_p^2}.
\end{equation}
The obtained dimensionless ratios $\hat \Pi_1$ and $\hat \Pi_2$ are geometrically small parameters, and thus  friction does not have a leading order effect on the system dynamics. For consistency, since terms of similar orders were previously neglected, $\hat \Pi_2$ is neglected hereafter. However, for some configurations $\hat \Pi_1\gg \hat \Pi_2$. Keeping $O(\hat\Pi_1)$ terms yields expressions for the fluid pressure ahead of the cylinder
\begin{equation}
P|_{X>\varepsilon_l L_p }=\frac{F_e-2\hat \Pi_1}{1+\hat \Pi_1}
\end{equation}
and the outer-region penetration length,
\begin{equation}
L_p=L_p(0)+\frac{F_e-2\hat\Pi_1}{1+\hat\Pi_1}.
\end{equation}

Thus, friction effects are of order of the small parameter $\hat\Pi_1$, and only slightly reduce the liquid pressure and the penetration length $L_p$. Following similar asymptotic matching procedure to that presented in \S5.1, the uniform solution for contact is given by
\begin{multline}
L_{p,\rm{uniform}}=\left(L_p(0)+\hat F_e\right)\Bigg\{1\\ + {\rm{W}}\left[ -{  \frac{1}{L_p(0)+\hat F_e}{\exp{\left(-1-\frac{ \hat T-(\hat F_e+L_p(0))\rm{Ln}\left(\hat F_e\right)-L_p(0)}{L_p(0)+\hat F_e} \right)}}} \right] \Bigg\},
\label{unifromLP}
\end{multline}
where $\hat F_e = ({F_e-2\hat\Pi_1})/({1+\hat\Pi_1})$ and $\hat T = t/t^*_{\textrm{shear}}$ (where $t^*_{\textrm{shear}}$ is defined in (\ref{tshear})). 

The uniform solution (\ref{unifromLP}) reaches a steady-state of $L_p\rightarrow L_p(0)+\hat F_e$. If $L_p(0)+\hat F_e<0$, the steady-state solution is not physical, indicating that the inner cylinder completely exits the tube. Thus, steady-state contact will lock the inner cylinder within the elastic tube only for the range of forces 
\begin{equation}
2\hat \Pi_1-L_p(0)(1+\hat \Pi_1)<F_e<-2
\end{equation}
Outside of this range the inner cylinder will exit the tube either due to fluid entering the tube $F_e>-2$ or sliding while in contact $F_e\leq 2\hat\Pi_1-L_p(0)(1+\hat\Pi_1)$. (In dimensional force, and omitting the small $O(\hat \Pi_1)$ terms, this range is given by $-\pi r_p^4Ew l_p(0)/2r_i^2 l<f_e<-2\pi Ewh_0 r_p^2/r_i^2$.)

The results presented in this subsection will be discussed further in the following section, summarizing the non-linear extraction section \S5.

\subsection{Summary of extraction dynamics}
\begin{figure} \centering
\includegraphics[width=0.9\textwidth]{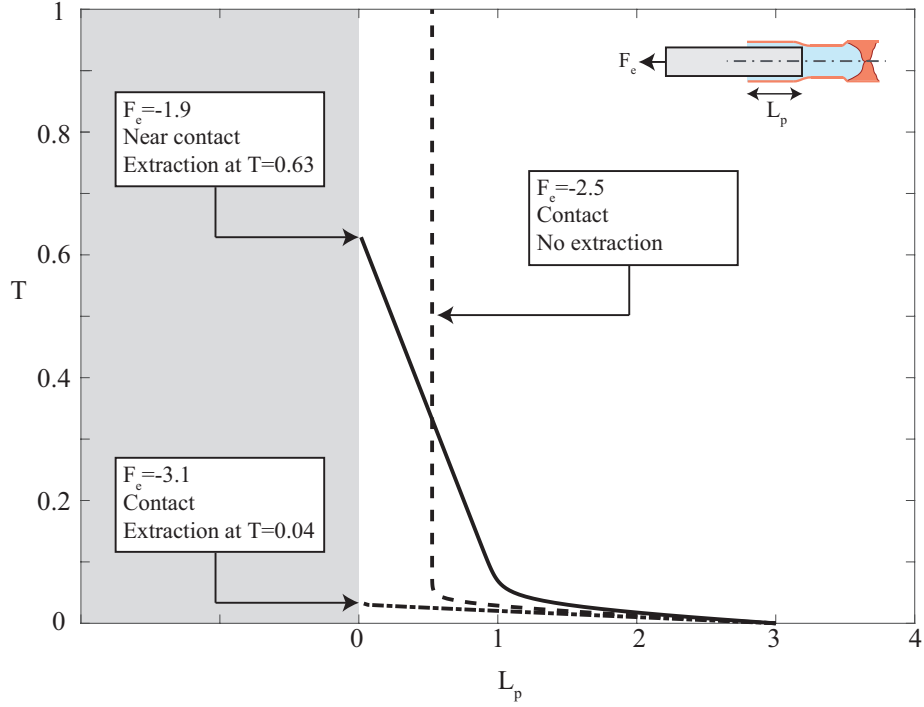}
\captionsetup{justification=justified}
\caption{\textbf{Extraction dynamics for near-contact (smooth line, $F_e>-2$), locked contact (dashed line, $2\tilde\Pi_1-L_p(0)(1+\tilde\Pi_1)<F_e<-2$) and extracted contact (dashed-dotted line, $F_e\leq 2\tilde\Pi_1-L_p(0)(1+\tilde\Pi_1)$) configurations}. For $F_e<-2$, the negative gauge pressure reduces, but not eliminates, the gap between the cylinder and the tube. Pressure-driven flow through this gap allows the gradual extraction of the inner cylinder. For $2\tilde\Pi_1-L_p(0)(1+\tilde\Pi_1)<F_e<-2$, after an initial transient motion, the cylinder is locked within the tube and negative gauge pressure creates contact between cylinder and the tube. Thus, there is no flow, and the inner cylinder is locked at a constant location in which there is a balance between the external force, friction, and the negative gauge pressure.  Finally, for $F_e\leq 2\tilde\Pi_1-L_p(0)(1+\tilde\Pi_1)$, the inner cylinder is extracted in the initial transient motion. For all presented cases $K=0.3$, $L_p(0)=3$, $\varepsilon_t=0.01$ and $\tilde \Pi_1=0.1$.}
\label{contact_SUM}
\end{figure}

Figure \ref{contact_SUM} summarizes the different dynamics obtained for the case of non-linear extraction. For near contact (smooth line, $F_e>-2$) the viscous resistance in the region near $\tilde Z=0$ is singular and determines the liquid mass flux, and thus the motion of the inner cylinder. After a shear-dominant early dynamics (which occurs for all cases), the cylinder moves in constant speed due to the steady conditions near $\tilde Z=0$. Increasing the extraction force to $2\tilde\Pi_1-L_p(0)(1+\tilde\Pi_1)<F_e<-2$ creates steady-state contact (dashed line) in which the extraction force is balanced with the force due to liquid pressure at a constant penetration length $L+p=L_p(0)+({F_e-2\hat\Pi_1})/({1+\hat\Pi_1})$. Friction creates only weak $O(\hat \Pi_1)$ effects. Extracting the tube with a greater force  $F_e\leq 2\tilde\Pi_1-L_p(0)(1+\tilde\Pi_1)$ completely removes the inner cylinder from the tube before a steady-state can be reached (dashed-dotted line), and thus loackage of the inner cylinder is limited to a specific range of extraction forces.

\section{Concluding remarks}
This work studied the low-Reynolds number fluid mechanics of a basic configuration, consisting of a slender cylinder inserted into a fluid-filled elastic tube, which is  relevant to various minimally invasive medical procedures. Governing integro-differential equations were derived by applying the lubrication approximation and thin shell elastic model. Solutions for various limits were obtained by  scaling analysis and regular and singular asymptotic schemes. Table 1 summarizes the different regions with regard to the value of $f_e$, the external force acting to extract or insert the cylinder to the tube. In addition, Table 1 presents comments and  descriptions on the dominant mechanisms in regions with no approximate solutions. 

\begin{table}
\begin{tabular}{l p{85mm}}
$f_e\gg E h_0w$	& Examined in \S4. Nonlinear dynamics involving three distinct regimes (with additional simplifying assumptions detailed in (\ref{ass2})). An early-time regime (\S4.2.2) governed by balance between shear stress, fluidic pressure and the external force. Intermediate regime (\S4.2.1) governed by the external force, fluid pressure and viscous-peeling of the inner cylinder from the external tube. Late-time regime (\S4.2.4) in which pressure-driven viscous flow exiting the tube determines the motion of the inner cylinder.\vspace{5px}  \\ \vspace{5px} 
$f_e\sim E h_0w$	& Not examined in this work. Nonlinear insertion dynamics create positive deformations which reduce viscous resistance, but do not involve a distinct propagation of a peeling front.\\ \vspace{5px} 
$|f_e|\ll E h_0w$	& Examined in \S3. Linearized dynamics representing a rigid configuration in leading order (see (\ref{(A)})), with small corrections due to elastic effects (see (\ref{UP300})). The small deformations create elastic potential energy, which may lead to motion in a direction opposite to the external transient force.   \\  \vspace{5px}
$ -\frac{2\pi Ew h_0r_p^2}{r_i^2}< f_e \sim -E h_0 w $	& Not examined in this work. Nonlinear extraction dynamics create negative deformations, thus increasing the viscous resistance. The front of the cylinder does not have a singular dominant effect on viscous resistance.  \\ \vspace{5px}
$f_e\rightarrow \left(-\frac{2\pi Ew h_0r_p^2}{r_i^2}\right)^+$	& Examined in \S5.1. Elastic deformation creates near contact between the tube and the elastic cylinder at $x=l_p$. The viscous resistance in the region near $x=l_p$ is singular and dominates the mass-flow outside of the  cylinder. After an early time region similar to \S4.2.2, the cylinder decelerates and exits the tube at a constant speed determined by the conditions near $x=l_p$ (see (\ref{unifromLP})). In this case the dynamics of the configuration are highly sensitive to the geometry at the tip of the penetrating cylinder. \\ \vspace{5px}
$-\frac{\pi r_p^4Ew l_p(0)}{2r_i^2 l}<f_e<-\frac{2\pi Ewh_0 r_p^2}{r_i^2}$	& Examined in \S5.2. In this case the extraction, and negative deformation, create contact between the inner cylinder and the elastic tube. After an early time region similar to  \S4.2.2, the cylinder decelerates and reaches a steady-state of balance between the external force, the fluid pressure and friction. In this range of force the inner cylinder remains at a constant position within the tube.\\ \vspace{5px}
$f_e <-\frac{\pi r_p^4Ew l_p(0)}{2r_i^2 l}$	& Examined in \S5.2. Similar to the previous case, however, for this range of extracting forces the cylinder is completely extracted from the tube before a steady-state balance is reached.
\end{tabular}
\caption{Summary of results for different values of $f_e$, the external insertion or extraction force.}
\end{table}

\acknowledgments{}

\bibliographystyle{jfm}
\bibliography{Bib_File}

\begin{thebibliography}{33}
\expandafter\ifx\csname natexlab\endcsname\relax\def\natexlab#1{#1}\fi

\bibitem[Abreu {\em et~al.\/}(2014)Abreu, Levant, Steinberg \&
  Seifert]{abreu2014fluid}
{\sc Abreu, David, Levant, Michael, Steinberg, Victor \& Seifert, Udo} 2014
  Fluid vesicles in flow. {\em Advances in colloid and interface science\/}
  {\bf 208}, 129--141.

\bibitem[Barenblatt(1952)]{barenblatt1952some}
{\sc Barenblatt, GI} 1952 On some unsteady motions of a liquid and gas in a
  porous medium. {\em Prikl. Mat. Mekh\/} {\bf 16}~(1), 67--78.

\bibitem[Camalet \& J{\"u}licher(2000)]{camalet2000generic}
{\sc Camalet, S{\'e}bastien \& J{\"u}licher, Frank} 2000 Generic aspects of
  axonemal beating. {\em New Journal of Physics\/} {\bf 2}~(1), 24.

\bibitem[Chew {\em et~al.\/}(1996)Chew, Buffington, Kendall, Osborn \&
  Woodsworth]{chew1996urethroscopy}
{\sc Chew, Dennis~J, Buffington, Tony, Kendall, Michael~S, Osborn, Steven~D \&
  Woodsworth, Bruce~E} 1996 Urethroscopy, cystoscopy, and biopsy of the feline
  lower urinary tract. {\em Veterinary Clinics: Small Animal Practice\/} {\bf
  26}~(3), 441--462.

\bibitem[Davis(2015)]{davis2015no}
{\sc Davis, Thomas} 2015 No-fluoroscopy crossing of chronic total occlusions
  using ocelot optical coherence tomography guided catheter. {\em Vascular
  Disease Management\/} {\bf 12}~(12), E230--E241.

\bibitem[Dunn \& Weisse(2015)]{dunn2015thrombectomy}
{\sc Dunn, Marilyn~E \& Weisse, Chick} 2015 Thrombectomy and thrombolysis: The
  interventional radiology approach. {\em Veterinary image-guided
  interventions\/} p. 464.

\bibitem[Duprat \& Stone(2015)]{duprat2015fluid}
{\sc Duprat, Camille \& Stone, Howard~A} 2015 {\em Fluid-Structure Interactions
  in Low-Reynolds-Number Flows\/}. Royal Society of Chemistry.

\bibitem[Elbaz \& Gat(2016)]{elbaz2016axial}
{\sc Elbaz, SB \& Gat, AD} 2016 Axial creeping flow in the gap between a rigid
  cylinder and a concentric elastic tube. {\em Journal of Fluid Mechanics\/}
  {\bf 806}, 580--602.

\bibitem[Heil(1996)]{heil1996stability}
{\sc Heil, Matthias} 1996 The stability of cylindrical shells conveying viscous
  flow. {\em Journal of Fluids and Structures\/} {\bf 10}~(2), 173--196.

\bibitem[Heil(1998)]{heil1998stokes}
{\sc Heil, Matthias} 1998 Stokes flow in an elastic tube - a large-displacement
  fluid-structure interaction problem. {\em International journal for numerical
  methods in fluids\/} {\bf 28}~(2), 243--265.

\bibitem[Hewitt {\em et~al.\/}(2015)Hewitt, Balmforth \&
  De~Bruyn]{hewitt2015elastic}
{\sc Hewitt, IJ, Balmforth, NJ \& De~Bruyn, JR} 2015 Elastic-plated gravity
  currents. {\em European Journal of Applied Mathematics\/} {\bf 26}~(1),
  1--31.

\bibitem[Karahalios(1990)]{karahalios1990some}
{\sc Karahalios, George~T} 1990 Some possible effects of a catheter on the
  arterial wall. {\em Medical Physics\/} {\bf 17}~(5), 922--925.

\bibitem[Kumar {\em et~al.\/}(2013)Kumar, Chandel, Kumar \& Kumar]{kumar}
{\sc Kumar, Harjeet, Chandel, RS, Kumar, Sanjeev \& Kumar, Sanjeet} 2013 A
  mathematical model for blood flow through a narrow catheterized artery. {\em
  International Journal of Theoretical \& Applied Sciences\/} {\bf 5}~(2),
  101--108.

\bibitem[Leal(2007)]{leal2007advanced}
{\sc Leal, L~Gary} 2007 {\em Advanced transport phenomena: fluid mechanics and
  convective transport processes\/}, , vol.~7. Cambridge University Press.

\bibitem[Lighthill(1968)]{lighthill1968pressure}
{\sc Lighthill, MJ} 1968 Pressure-forcing of tightly fitting pellets along
  fluid-filled elastic tubes. {\em Journal of Fluid Mechanics\/} {\bf 34}~(1),
  113--143.

\bibitem[Lister {\em et~al.\/}(2013)Lister, Peng \& Neufeld]{lister2013viscous}
{\sc Lister, John~R, Peng, Gunnar~G \& Neufeld, Jerome~A} 2013 Viscous control
  of peeling an elastic sheet by bending and pulling. {\em Physical review
  letters\/} {\bf 111}~(15), 154501.

\bibitem[Marzo {\em et~al.\/}(2005)Marzo, Luo \& Bertram]{marzo2005three}
{\sc Marzo, A, Luo, XY \& Bertram, CD} 2005 Three-dimensional collapse and
  steady flow in thick-walled flexible tubes. {\em Journal of Fluids and
  Structures\/} {\bf 20}~(6), 817--835.

\bibitem[Nacey \& Delahijnt(1993)]{nacey1993evolution}
{\sc Nacey, John \& Delahijnt, Brett} 1993 The evolution and development of the
  urinary catheter. {\em Australian and New Zealand Journal of Surgery\/} {\bf
  63}~(10), 815--819.

\bibitem[Park {\em et~al.\/}(2018)Park, Tixier, Christensen, Arnbjerg-Nielsen,
  Zwieniecki \& Jensen]{park2018viscous}
{\sc Park, Keunhwan, Tixier, Aude, Christensen, AH, Arnbjerg-Nielsen, SF,
  Zwieniecki, MA \& Jensen, KH} 2018 Viscous flow in a soft valve. {\em Journal
  of Fluid Mechanics\/} {\bf 836}.

\bibitem[Pellerin {\em et~al.\/}(2014)Pellerin, Maleux, D{\'e}an, Pernot,
  Golzarian \& Sapoval]{pellerin2014microvascular}
{\sc Pellerin, Olivier, Maleux, Geert, D{\'e}an, Carole, Pernot, Simon,
  Golzarian, Jafar \& Sapoval, Marc} 2014 Microvascular plug: a new embolic
  material for hepatic arterial skeletonization. {\em Cardiovascular and
  interventional radiology\/} {\bf 37}~(6), 1597--1601.

\bibitem[Rogers \& Laird(2007)]{rogers2007overview}
{\sc Rogers, Jason~H \& Laird, John~R} 2007 Overview of new technologies for
  lower extremity revascularization. {\em Circulation\/} {\bf 116}~(18),
  2072--2085.

\bibitem[Sarkar \& Jayaraman(2001)]{sarkar2001nonlinear}
{\sc Sarkar, A \& Jayaraman, G} 2001 Nonlinear analysis of oscillatory flow in
  the annulus of an elastic tube: application to catheterized artery. {\em
  Physics of Fluids\/} {\bf 13}~(10), 2901--2911.

\bibitem[Serruys {\em et~al.\/}(1993)Serruys, Foley \&
  De~Feyter]{serruys1993quantitative}
{\sc Serruys, Patrick~W, Foley, David~P \& De~Feyter, Pim~J} 1993 {\em
  Quantitative coronary angiography in clinical practice\/}, , vol. 145.
  Springer Science \& Business Media.

\bibitem[Tani {\em et~al.\/}(2017)Tani, Cambau, Bico \&
  Reyssat]{tani2017motion}
{\sc Tani, Marie, Cambau, Thomas, Bico, Jose \& Reyssat, Etienne} 2017 Motion
  of a rigid sphere through an elastic tube with a lubrication film. In {\em
  APS Meeting Abstracts\/}.

\bibitem[Timoshenko \& Woinowsky-Krieger(1959)]{timoshenko1959theory}
{\sc Timoshenko, Stephen~P \& Woinowsky-Krieger, Sergius} 1959 {\em Theory of
  plates and shells\/}. McGraw-hill.

\bibitem[T{\"o}zeren {\em et~al.\/}(1982)T{\"o}zeren, {\"O}zkaya \&
  T{\"o}zeren]{tozeren1982flow}
{\sc T{\"o}zeren, Aydin, {\"O}zkaya, Nihat \& T{\"o}zeren, H{\"u}sn{\"u}} 1982
  Flow of particles along a deformable tube. {\em Journal of biomechanics\/}
  {\bf 15}~(7), 517--527.

\bibitem[Vajravelu {\em et~al.\/}(2011)Vajravelu, Sreenadh, Devaki \&
  Prasad]{vajravelu2011mathematical}
{\sc Vajravelu, Kuppalapalle, Sreenadh, Sreedharamalle, Devaki, Palluru \&
  Prasad, Kerehalli} 2011 Mathematical model for a herschel-bulkley fluid flow
  in an elastic tube. {\em Open Physics\/} {\bf 9}~(5), 1357--1365.

\bibitem[V{\'a}zquez(2007)]{vazquez2007porous}
{\sc V{\'a}zquez, Juan~Luis} 2007 {\em The porous medium equation: mathematical
  theory\/}. Oxford University Press.

\bibitem[Vlahovska {\em et~al.\/}(2011)Vlahovska, Young, Danker \&
  Misbah]{vlahovska2011dynamics}
{\sc Vlahovska, PM, Young, Y-N, Danker, G \& Misbah, C} 2011 Dynamics of a
  non-spherical microcapsule with incompressible interface in shear flow. {\em
  Journal of Fluid Mechanics\/} {\bf 678}, 221--247.

\bibitem[Weisstein(2002)]{weisstein2002lambert}
{\sc Weisstein, EW} 2002 Lambert w-function.

\bibitem[Wiggins \& Goldstein(1998)]{wiggins1998flexive}
{\sc Wiggins, Chris~H \& Goldstein, Raymond~E} 1998 Flexive and propulsive
  dynamics of elastica at low reynolds number. {\em Physical Review Letters\/}
  {\bf 80}~(17), 3879.

\bibitem[Zaitsev \& Polyanin(2002)]{zaitsev2002handbook}
{\sc Zaitsev, VF \& Polyanin, AD} 2002 {\em Handbook of exact solutions for
  ordinary differential equations\/}. CRC press.

\bibitem[Zel'dovich \& Kompaneets(1950)]{zel1950towards}
{\sc Zel'dovich, YB \& Kompaneets, AS} 1950 Towards a theory of heat conduction
  with thermal conductivity depending on the temperature. {\em Collection of
  papers dedicated to 70th birthday of Academician AF Ioffe, Izd. Akad. Nauk
  SSSR, Moscow\/} pp. 61--71.

\end{thebibliography}

\end{document}